\documentclass[conference]{IEEEtran}
\IEEEoverridecommandlockouts
\usepackage{cite}
\usepackage{svg}
\usepackage{amsmath,amssymb,amsfonts}
\usepackage{algorithmic}
\usepackage{graphicx}
\usepackage{textcomp}
\usepackage{xcolor}
\usepackage{booktabs} 
\usepackage{graphicx}
\usepackage{xcolor}
\usepackage{xspace}
\usepackage[normalem]{ulem}
\useunder{\uline}{\ul}{}
\usepackage{lscape}
\usepackage{rotating}
\usepackage{supertabular}
\usepackage{longtable}
\usepackage{enumerate}
\usepackage{svg}
\usepackage{amsmath}
\usepackage{colortbl}
\usepackage{subcaption}
\usepackage{hhline}
\usepackage{makecell}
\usepackage{caption}
\usepackage{hyperref}
\usepackage{enumitem}
\usepackage{tcolorbox}
\usepackage{tabularx}
\usepackage{lipsum}
\usepackage{listings}
\usepackage{tabularx,ragged2e}
\usepackage{verbatim}
\usepackage{fancyvrb}
\usepackage{balance}
\usepackage{multirow}

\def\BibTeX{{\rm B\kern-.05em{\sc i\kern-.025em b}\kern-.08em
    T\kern-.1667em\lower.7ex\hbox{E}\kern-.125emX}}
\begin{document}

\title{Decisions in Continuous Integration and Delivery: An Exploratory Study
\thanks{This work is supported by National Key R\&D Program of China with No. 2018YFB1402800, NSFC with No. 62172311, Hubei Provincial Natural Science Foundation of China with No. 2021CFB577, and Research Foundation of Shenzhen Polytechnic with No. 6022312043K.}
}




\author{
    \IEEEauthorblockN{Yajing Luo\IEEEauthorrefmark{2}, Peng Liang\IEEEauthorrefmark{2}\IEEEauthorrefmark{1}, Mojtaba Shahin\IEEEauthorrefmark{3}, Zengyang Li\IEEEauthorrefmark{4}, Chen Yang\IEEEauthorrefmark{5}\thanks{DOI reference number: 10.18293/SEKE2022-171}}
    \IEEEauthorblockA{
    \IEEEauthorrefmark{2}School of Computer Science, Wuhan University, Wuhan, China\\
    \IEEEauthorrefmark{3}School of Computing Technologies, RMIT University, Melbourne, Australia\\
    \IEEEauthorrefmark{4}School of Computer Science \& Hubei Provincial Key Laboratory of Artificial Intelligence and Smart Learning, \\Central China Normal University, Wuhan, China \\
    \IEEEauthorrefmark{5}School of Artificial Intelligence, Shenzhen Polytechnic, Shenzhen, China\\
    \{luoyajing, liangp\}whu.edu.cn, mojtaba.shahin@rmit.edu.au, zengyangli@ccnu.edu.cn, yangchen@szpt.edu.cn}
}

\maketitle

\begin{abstract}
In recent years, Continuous Integration (CI) and Continuous Delivery (CD) has been heatedly discussed and widely used in part or all of the software development life cycle as the practices and pipeline to deliver software products in an efficient way. There are many tools, such as Travis CI, that offer various features to support the CI/CD pipeline, but there is a lack of understanding about what decisions are frequently made in CI/CD. In this work, we explored one popular open-source project on GitHub, Budibase, to provide insights on the types of decisions made in CI/CD from a practitioners’ perspective. We first explored the GitHub Trending page, conducted a pilot repository extraction, and identified the Budibase repository as the case for our study. We then crawled all the closed issues from the repository and got 1,168 closed issues. Irrelevant issues were filtered out based on certain criteria, and 370 candidate issues that contain decisions were obtained for data extraction. We analyzed the issues using a hybrid approach combining pre-defined types and the Constant Comparison method to get the categories of decisions. The results show that the major type of decisions in the Budibase closed issues is \emph{Functional Requirement Decision} (67.6\%), followed by \emph{Architecture Decision} (11.1\%). Our findings encourage developers to put more effort on the issues and making decisions related to CI/CD, and provide researchers with a reference of decision classification made in CI/CD.
\end{abstract}


\section{Introduction}\label{introduction}

The Software Development Life Cycle (SDLC) of an information system goes through the steps of planning, creation, testing, and deployment. In the past, organizations needed to provision, operate, and maintain the build job creation, processing, and reporting themselves \cite{Gallaba2020Use}. In recent years, DevOps, a set of practices (e.g., cloud-based continuous integration, and automated deployment) that combine software development and IT operations, enables organizations to deliver changes into production as quickly as possible, without compromising software quality \cite{Zhao2017The}.
\textbf{In this study}, we focus on Continuous Integration and Delivery (CI/CD), the key enabler of DevOps. CI systems automate the compilation, building, and testing of software \cite{Hilton2016Usage}. A typical CI service is composed of three types of nodes. First, build job creation nodes queue up new build jobs when configured build events occur. Next, a set of build job processing nodes process build jobs from the queue, adding job results to another queue. Finally, build job reporting nodes process job results, updating team members of the build status \cite{Gallaba2020Use}. CD aims at ensuring an application is always at production-ready state after successfully passing automated tests and quality checks \cite{Shahin2017Continuous}.

Nowadays, there are many tools that have been available to support the CI/CD pipeline, such as Jenkins\footnote{\url{https://www.jenkins.io/}}, Travis CI\footnote{\url{https://travis-ci.org/}}, Circle CI\footnote{\url{https://circleci.com/}}, and GitHub Actions\footnote{\url{https://github.com/features/actions}}, offering various CI/CD features. Among them, GitHub Actions is a CI/CD platform that allows developers to automate their build, test, and deployment pipeline. It was launched in 2018 and supported CI/CD from November 13, 2019. With GitHub Actions, practitioners can create workflows that build and test every pull request to their repository, or deploy merged pull requests to production.

Although many studies have been conducted in a CI/CD context \cite{Gallaba2020Use}\cite{Shahin2019An}, very few of them have explored decisions made in CI/CD, how the decisions made in CI/CD differ from those made in traditional software development, and what types of decisions are made in CI/CD. We selected the Budibase repository as the case for this study due to: (1) it uses GitHub Actions to support CI/CD, (2) it is a trending repository on GitHub in the recent three months, (3) it is active in the recent three months, and (4) the number of its closed issues exceeds 1000. In short, we selected the Budibase repository as the study project and collected data from its closed issues. We then identified and extracted decisions from the dataset, and classified them using a hybrid approach.

\textbf{The contributions of this paper}: (1) we provide an exploratory study on the decisions made in CI/CD; (2) we classify the decisions made in CI/CD; and (3) we compare the decisions made in CI/CD with the decisions made in traditional development.

The rest of the paper is organized as follows: Section \ref{related} discusses related work to our study. Section \ref{research} describes the research question and its rationale, the data collection, filtering, extraction, and analysis process. Section \ref{res} provides the results of the research question. Section \ref{dis} explains the results with their implications and comparison with the existing work. Section \ref{threat} presents the threats to the validity of this study. Section \ref{conc} concludes this work with further directions.

\section{Related Work}\label{related}

\subsection{Continuous Integration and Delivery}
\label{sec:CI/CD}
CI is an established software quality assurance practice and has been the focus of much prior research with a diverse range of methods and populations \cite{Widder2019A}. CD has been adopted by many software organizations to develop and deliver quality software more frequently and reliably \cite{Shahin2019An}.

Gallaba \textit{et al.} \cite{Gallaba2020Use} set out to study how CI features are being used and misused in 9,312 open source systems that use Travis CI. Widder \textit{et al.} \cite{Widder2019A} reviewed the CI literature of 37 papers from the perspective of pain points to adoption and usage, and performed a mixed-methods conceptual replication of previously observed findings. Zhao \textit{et al.} \cite{Zhao2017The} studied the adaptation and evolution of code writing and submission, issue and pull request closing, and testing practices on hundreds of established projects on GitHub adopting Travis CI. Hilton \textit{et al.} \cite{Hilton2016Usage} used three complementary methods to study the usage of CI in open-source projects. As for CD, Shahin \textit{et al.} \cite{Shahin2019An} conducted a mixed-methods empirical study and presented a conceptual framework to support the process of (re-)architecting for CD.

To the best of our knowledge, there are no studies that explore the decisions made in CI/CD, how they differ from those in traditional software development, and what types of decisions are frequently made. In our study, we intended to explore the decisions made in CI/CD using an open source project (Budibase) that employs GitHub Actions.

\subsection{Decisions in Software Engineering}
\label{sec:Decisions}

Developing and maintaining a software system involves making numerous important decisions by stakeholders \cite{Shahbazian2018Recovering}. These decisions cover the life cycle of software development, including requirements~\cite{olsson2019empirical}, architecture design~\cite{shahin2009architectural}, project management~\cite{drury2014performance}, etc. \cite{Li2019Decisions}, and each decision is typically influenced by other decisions and involves trade-offs in system properties \cite{Shahbazian2018Making}. Design decisions directly impact system quality \cite{Shahbazian2018Making}, and high-quality decisions are critical to the success of projects \cite{PMI2015Capturing}. Many researchers focus on exploring how to make high-quality and appropriate decisions that meet project objectives and maximize system benefits, analyzing the rationale behind decisions, and understanding their effects on the system.

Tang \textit{et al.} \cite{Tang2021Decision-Making} proposed a systematic approach to software design decision-making. They broke decision-making down into nine principles that can be taught, learned, and practiced, and each principle addresses one decision-making aspect that focuses on a specific type of information used in making and evaluating decisions. Shahbazian \textit{et al.} \cite{Shahbazian2018Making} posed the problem of making complex, interacting design decisions relatively early in the project’s lifecycle and outlined a search-based and simulation-based approach for helping architects make these decisions and understand their effects. Li \textit{et al.} \cite{Li2019Decisions} tried to identify decisions discussed in the Hibernate developer mailing list and explore decision-making from several aspects (i.e., description, classification, underlying rationale, supporting approaches, related artifacts, and trend). Sharma \textit{et al.} \cite{Sharma2021Extracting} explored the rationale behind ‘how’ the decisions are made, and made a methodological contribution by presenting a heuristics-based rationale extraction system employing multiple heuristics, and following a data-driven, bottom-up approach to infer the rationale behind specific decisions. Liu \textit{et al.} \cite{liu2019understanding} intended to understand how decisions are made in requirements engineering through the student projects in a requirements engineering course.

Different from the above works, we studied the decisions made in CI/CD by analyzing the closed issues in the Budibase repository and focusing on the types of decisions.

\section{Research Design}\label{research}

We set the goal of this study based on the Goal-Question-Metric approach \cite{Basili1994TheGQ}, which is to \textbf{analyze} the decisions in CI/CD \textbf{for the purpose of} characterizing \textbf{with respect to} the types of decisions made in CI/CD \textbf{from the point of view of} practitioners \textbf{in the context of} CI/CD of open source software. To archive this goal, we address the following key Research Question (RQ):

\begin{itemize}
\item
  \textbf{RQ}: What types of decisions are made in CI/CD?
  
  \textbf{Rationale}: Various types of decisions made in traditional software development have been discussed in literature (e.g.,  \cite{Li2019Decisions}). This RQ aims to provide a classification of the decisions made in projects using CI/CD and compare these decisions with those made in traditional software development, so that we can see the difference between CI/CD and traditional development from a decision perspective.
\end{itemize}

\begin{figure*}[h]
	\centering
	\includegraphics[width=0.65\linewidth]{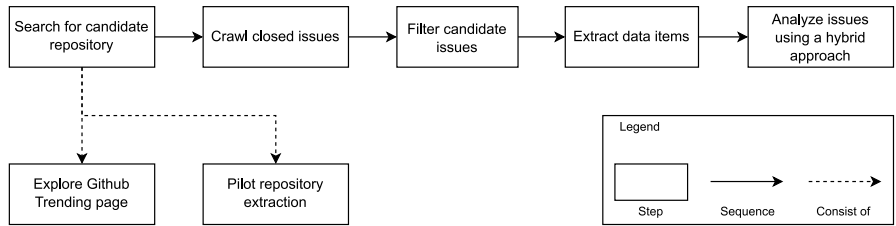}
	\caption{Overview of the research process}
	\label{fig:Overview of research process}
\end{figure*}

In the following subsections, we detail the research process (see Figure \ref{fig:Overview of research process}) used to answer this RQ.

\subsection{Data Collection}
\label{dataCollection}
To answer the RQ, we need to identify and collect decisions from the projects using CI/CD. GitHub is a popular and most frequently used code hosting platform for version control and collaboration. After exploring several artifacts (i.e., Github Discussions, Github Issues, Slack, Gitter, and Microsoft teams), we intended to select a suitable GitHub project that employs CI/CD practices for our study, and use GitHub Issues (developer discussions) as the data source, since closed issues include a wide range of decisions made in development. We searched for the candidate GitHub project based on the following criteria:

\begin{itemize}
\item
  \textbf{A repository running a GitHub Actions workflow}

Given that we needed to select a project applying CI/CD. GitHub provides a CI/CD tool named GitHub Actions to allow developers to build, test, and deploy their software automatically. Github Actions provided by Github started to support CI/CD in 2019 and does not require migration from Circle CI, Jenkins, or Travis CI. We believe that there should be many projects on Github applying Github Actions in the last 2 years. Hence, we decided to look for a repository on GitHub that uses the feature of GitHub Actions.
\item
  \textbf{A hot topic in GitHub in the last three months}

  A GitHub Trending page is available on GitHub where trending repositories are listed based on the number of times they have been
  starred by users every day, week, or month. This trending page provides a view of the open-source projects which the community is most excited about. We decided to trace the hot topics in GitHub in the last three months and select an appropriate repository for our study.
\item
  \textbf{An active repository}

  We decided to choose a repository that is currently active and frequently updated (e.g., usually make changes to a file and push them to GitHub as commits, often open and merge a pull request, and have heated discussion in GitHub Issues).
\end{itemize}

\begin{itemize}
\item
  \textbf{The number of closed issues exceeds 1000}

  Since issues are the data source of this study, and the number of issues of a project is an indicator of the project size, which is an important factor to get enough data for an empirical study. We decided to look for a repository with more than 1000 closed issues.
\end{itemize}

After exploring the repositories listed on the GitHub Trending page for the past three months, we picked out several repositories (including Budibase, Rails, and Tabby), which met the above criteria. We then conducted a pilot repository selection. The first author randomly selected 50 issues from the closed issues of the candidate repositories. The purpose was to verify whether the repositories we picked out were appropriate and whether the data item (type of decision) can be extracted from the issues (see the criteria in Section \ref{postclassproc}). Next, the first author extracted the data item from these 50 issues and then discussed the extraction results with the second author to see if the extraction results were correct. After the pilot repository selection, we finally chose Budibase\footnote{\url{https://github.com/Budibase/budibase}}.

Budibase is an all-in-one low-code platform for building, designing, and automating business apps with which building business apps in minutes. We crawled all the URLs of the closed issues of Budibase into an Excel sheet on January, 2022, and got 1,168 links.

\subsection{Data Filtering, Extraction and Analysis}
\label{postclassproc}

\subsubsection{Filter candidate issues}
The criteria for filtering issues are defined as follows:

\begin{itemize}
\item
  \textbf{If an issue contains the data item (i.e., type of decision) to be extracted, we include it.}
\item
  \textbf{If an issue does not contain the data item to be extracted or cannot be understood, we exclude it.}
\end{itemize}

This step was conducted by the first author, and the issues that the first author could not decide were discussed among the authors to reach a consensus. After excluding the irrelevant issues, we finally got 370 issues out of the 1,168 closed issues from Budibase.

\subsubsection{Extract and analyze the data item}
First of all, for answering the RQ, the first author extracted the data item (type of decision) from each relevant issue. After data extraction, the first author classified the decisions extracted based on the decision types provided by Li \textit{et al.} \cite{Li2019Decisions}. Then the first author discussed the classification results (i.e., by following the pre-defined types in \cite{Li2019Decisions}) with the second author and adjusted the classification by taking into account both the content of the extracted decisions and the labels of the issues using the Constant Comparison method \cite{Glaser1965The}. We refer to this combination of the pre-defined classification and the Constant Comparison method as a hybrid approach. We obtained the final classification of decisions using the hybrid approach. We have also provided the dataset and the classification results of decisions online \cite{replpack}.


\begin{table*}[h]
\caption{A classification of decisions made in the Budibase closed issues}
\label{A Classification of decisions in the Budibase closed issues}
\resizebox{\linewidth}{!}{
\begin{tabular}{|ll|l|l|c|}
\hline
\multicolumn{2}{|l|}{\textbf{Type}}                                                                                                                                                                & \textbf{Description}                                                                                                                                                                                                                                                                                                                                                                   & \textbf{Example}                                                                                                                                                                                                                                                                                                                  & \multicolumn{1}{l|}{\textbf{Percentage}} \\ \hline

\multicolumn{1}{|l|}{}                                                                                  & \begin{tabular}[c]{@{}l@{}}Functional \\ Requirement \\ Decision\end{tabular}            & { \begin{tabular}[c]{@{}l@{}}A description of the decisions on services that a software \\ system is supposed to accomplish involving calculations, \\ technical details, data manipulation and processing, and \\ other specific functionality.\end{tabular}}                                                                                                     & \textit{\begin{tabular}[c]{@{}l@{}}As a user, I want to be able to configure more advanced \\ searching capabilities on the search component, so that \\ I can find what I am looking for with a higher degree of \\ accuracy.\end{tabular}}                                                                                      & 67.6\%                                   \\ \cline{2-5} 
\multicolumn{1}{|l|}{\multirow{-2}{*}{\begin{tabular}[c]{@{}l@{}}Requirement \\ Decision\end{tabular}}} & \begin{tabular}[c]{@{}l@{}}Non-functional \\ Requirement \\ Decision\end{tabular}        & \begin{tabular}[c]{@{}l@{}}A description of the quality attributes of a software \\ system, judging the software system based on non-functional \\ standards that are critical to the success of the software \\ system.\end{tabular}                                                                                                                                                  & \textit{\begin{tabular}[c]{@{}l@{}}When I'm setting up a relationship in budibase, the \\ current relationship configuration UI is confusing and \\ difficult to remember how to use, I want to have a \\ simpler and more consistent experience with internal \\ relationships so it's easier to set up.\end{tabular}}           & 5.4\%                                    \\ \hline

\multicolumn{2}{|l|}{Architecture Decision}                                                                                                                                                        & {\begin{tabular}[c]{@{}l@{}}A description of the decisions with regard to architectural \\ additions, subtractions, and modifications to the software \\ architecture, the rationale, and the design rules, design \\ constraints, and additional requirements that (partially) \\ realize one or more requirements on a given architecture \cite{Jansen2005Software}.\end{tabular}} & \textit{\begin{tabular}[c]{@{}l@{}}Replace AppImage with Snap and RPM installations for \\ linux.\end{tabular}}                                                                                                                                                                                                                   & 11.1\%                                   \\ \hline

\multicolumn{1}{|l|}{}                                                                                  & \begin{tabular}[c]{@{}l@{}}Version \\ Control \\ Decision\end{tabular}                   & \begin{tabular}[c]{@{}l@{}}A description of the decisions on tracking and managing \\ changes to source code in version control systems.\end{tabular}                                                                                                                                                                                                                                  & \textit{\begin{tabular}[c]{@{}l@{}}I noticed in the input component that it was still using \\ the old css vars, these need to be updated.\end{tabular}}                                                                                                                                                                          & 1.6\%                                    \\ \cline{2-5} 
\multicolumn{1}{|l|}{\multirow{-2}{*}{\begin{tabular}[c]{@{}l@{}}Management \\ Decision\end{tabular}}}  & \begin{tabular}[c]{@{}l@{}}Documentation \\ Decision\end{tabular}                        & \begin{tabular}[c]{@{}l@{}}A description of the decisions related to documentation, \\ e.g., changes to README files, docs, and GitHub discussions,\\  etc.\end{tabular}                                                                                                                                                                                                               & \textit{\begin{tabular}[c]{@{}l@{}}Stage 1 Going to add code documentation of all functions, \\ in core, server and builder. Stage 2 (maybe a future \\ issue) Use document code to produce markdown files, and \\ publish to GitBook @ apidocs.budibase.com (does not exist \\ yet).\end{tabular}}                               & 3.2\%                                    \\ \hline
\multicolumn{1}{|l|}{}                                                                                  & \begin{tabular}[c]{@{}l@{}}Source Code \\ Decision\end{tabular}                          & \begin{tabular}[c]{@{}l@{}}A description of the decisions on making changes to source \\ code.\end{tabular}                                                                                                                                                                                                                                                                            & \textit{\begin{tabular}[c]{@{}l@{}}Create a new Options type in the backend/constants file. \\ Remove the Categories values list from the string type in \\ CreateEditColumn. Account for the new options field \\ everywhere that we are checking for \\ field.constraints.inclusion.\end{tabular}}                              & 2.2\%                                    \\ \cline{2-5} 
\multicolumn{1}{|l|}{\multirow{-2}{*}{\begin{tabular}[c]{@{}l@{}}Build \\ Decision\end{tabular}}}       & \begin{tabular}[c]{@{}l@{}}Continuous \\ Integration \\ Decision\end{tabular}            & \begin{tabular}[c]{@{}l@{}}A description of the decisions concerning the compilation \\ and building in CI pipeline supported development.\end{tabular}                                                                                                                                                                                                                                & \textit{\begin{tabular}[c]{@{}l@{}}Have a rollback mechanism on that CI job, so that we can \\ roll back to the previous release, or a pre-specified \\ version.\end{tabular}}                                                                                                                                                    & 0.3\%                                    \\ \hline
\multicolumn{1}{|l|}{}                                                                                  & \begin{tabular}[c]{@{}l@{}}Bug Fixing \\ Decision\end{tabular}                           & \begin{tabular}[c]{@{}l@{}}A description of the decisions made to fix the problem when \\ a bug is generated.\end{tabular}                                                                                                                                                                                                                                                             & \textit{\begin{tabular}[c]{@{}l@{}}This fix for this is simply parsing the data in data \\ providers for any datetime fields and building the lucene \\ query with ISO strings like before.\end{tabular}}                                                                                                                         & 4.6\%                                    \\ \cline{2-5} 
\multicolumn{1}{|l|}{}                                                                                  & \begin{tabular}[c]{@{}l@{}}Traditional \\ Testing \\ Decision\end{tabular}               & \begin{tabular}[c]{@{}l@{}}A description of the decisions on examining the artifacts \\ and the behavior of a software system under test by \\ validation and verification.\end{tabular}                                                                                                                                                                                            & \textit{\begin{tabular}[c]{@{}l@{}}When testing automations on the platform, I want to be \\ able to see the data flow from one block to the other so \\ that I can identify the inputs/outputs of each block and \\ debug with more ease.\end{tabular}}                                                                          & 4.3\%                                    \\ \cline{2-5} 
\multicolumn{1}{|l|}{\multirow{-3}{*}{\begin{tabular}[c]{@{}l@{}}Testing \\ Decision\end{tabular}}}     & \begin{tabular}[c]{@{}l@{}}Continuous \\ Integration \\ Testing \\ Decision\end{tabular} & \begin{tabular}[c]{@{}l@{}}A description of the decisions with respect to the testing \\ in CI pipeline supported development.\end{tabular}                                                                                                                                                                                                                                            & \textit{\begin{tabular}[c]{@{}l@{}}Individual unit tests for each automation block. Ability \\ to test automation integration, i.e., provide automation \\ schema and make sure it runs successfully.\end{tabular}}                                                                                                               & 0.3\%                                    \\ \hline
\multicolumn{1}{|l|}{}                                                                                  & \begin{tabular}[c]{@{}l@{}}Traditional \\ Deployment \\ Decision\end{tabular}            & \begin{tabular}[c]{@{}l@{}}A description of the decisions regarding all the activities \\ that make a software system available for use \cite{Pressman2014Software}.\end{tabular}                                                                                                                                                                                                                 & \textit{\begin{tabular}[c]{@{}l@{}}When a user deploys a budibase application, we need to \\ update data correctly in our dynamoDB database so that \\ users deployment quotas are managed correctly, and that \\ deployments will fail if they are going to go over the \\ current quota for a particular account.\end{tabular}} & 4.3\%                                    \\ \cline{2-5} 
\multicolumn{1}{|l|}{\multirow{-2}{*}{\begin{tabular}[c]{@{}l@{}}Deployment \\ Decision\end{tabular}}}  & \begin{tabular}[c]{@{}l@{}}Continuous \\ Deployment \\ Decision\end{tabular}              & \begin{tabular}[c]{@{}l@{}}A description of the decisions made when code changes are \\ automatically deployed to a production environment through \\ a pipeline as soon as they are ready, without human \\ intervention \cite{Skelton2016Continuous}.\end{tabular}                                                                                                                                      & \textit{\begin{tabular}[c]{@{}l@{}}Automated CI pipeline to deploy the latest helm chart to \\ the pre-prod environment.\end{tabular}}                                                                                                                                                                                            & 1.4\%                                    \\ \hline
\end{tabular}
}
\end{table*}

\section{Results}\label{res}

We applied the hybrid approach and classified the decisions into several types as shown in Table \ref{A Classification of decisions in the Budibase closed issues}, which provides the types (categories and subcategories) with their descriptions, examples, and percentages.

We got six categories of decisions made in CI/CD, namely \emph{Requirement Decision}, \emph{Architecture Decision}, \emph{Management Decision}, \emph{Build Decision}, \emph{Testing Decision}, and \emph{Deployment Decision}. We also got 11 subcategories of these major categories except \emph{Architecture Decision}, which does not have any subcategories. The major type of decisions in the Budibase closed issues is \emph{Functional Requirement Decision} (67.6\%), followed by \emph{Architecture Decision} (11.1\%). The rest types of decisions are all below 10\%. 5.4\% of the decisions belong to \emph{Non-functional Requirement Decision} while 4.6\% of the decisions are \emph{Bug Fixing Decision}. The percentages of \emph{Traditional Testing Decision} (4.3\%) and \emph{Traditional Deployment Decision} (4.3\%) are the same. The percentages of \emph{Continuous Integration Decision} (0.3\%) and \emph{Continuous Integration Testing Decision} (0.3\%) are the least. Note that, since one issue may contain multiple types of decisions, the sum of the percentages of all types of decisions is greater than 100\%.

\section{Discussion}\label{dis}
We first explain the results of this study and discuss their implications for practitioners and researchers. We then compare the study results, i.e., decisions made in CI/CD, with the decisions made in traditional open source development without using CI/CD. 

\subsection{Types of Decisions Made in CI/CD}
\subsubsection{Interpretations}
The Software Development Life Cycle (SDLC) simply outlines the tasks required to put together a software application. In the planning phase, requirements are defined to determine what the application is supposed to do and what quality attributes need to be met. Therefore, it can be seen from the statistics that, 67.6\% of the decisions are for functional requirements and 5.4\% are for non-functional requirements, which is reasonable that functional requirements are the major part to make a decision in requirements analysis. Then the creation phase models the way a software application will work. Architecture design is part of the outcome of this phase, and our result shows that 11.1\% of the issues contain architecture decisions. Based on the architecture design, developers start the actual implementation of the application. Usually, an open source project is implemented by a virtual team and the implementation tasks can be broken down into jobs for each developer, and consequently a source code management tool is used to help developers track changes to the code. We found that 1.6\% of the issues contain version control decisions. During the coding process, developers do not just code, they also write instructions and explanations about the code and developed application. Documentation, such as user guides, is written to give users a quick tour of the application's basic features, while comments in the source code provide further information about the code for other developers. In our study, we found that 3.2\% of the decisions are related to documentation. Before making an application available to users, it is critical to test it, and 4.3\% of the traditional testing decisions are identified in the result. When an error occurs, we need to decide how to fix it, and we can see 4.6\% of the decisions belonging to bug fixing. In the deployment phase, the application is made available in the user or production environment. 4.3\% of the issues are relevant to traditional deployment decisions. In the CI/CD pipeline, one advantage is that some tasks in SDLC, especially integration and deployment, can be automated. Our statistical result shows that 0.3\% of the issues belong to continuous integration decisions, 0.3\% of the issues contain continuous integration testing decisions, and 1.4\% of the decisions are about continuous deployment.

As we can observe from the results, a large percentage of decisions are made during the planning and creation phase of software development, followed by bug fixing and testing phrase. Given that the subject of the study (the selected project Budibase) is a CI/CD pipeline supported development project, we only got a small number of CI/CD-related decisions (e.g., \emph{Continuous Deployment Decision} and \emph{Continuous Integration Decision}). The possible reasons of this finding are that developers in CI/CD still focus on the traditional decision types and CI/CD only provides support to the project (e.g., by CI/CD pipeline tools) without the necessity of making many CI/CD-related decisions, or CI/CD-related decisions are not discussed and made in GitHub Issues, but in GitHub Actions, which requires further investigation.

\subsubsection{Implications}
For practitioners, our result suggests that they should pay more attention to the management, build, testing, and deployment decisions as they are as equally important as requirements and architecture decision in CI/CD. Practitioners should also put more effort on the issues and making decisions related to CI/CD, such as \emph{Continuous Integration Decision}, which are fundamental decisions to facilitate the CI/CD pipeline.

For researchers, we provide a dataset of various types of decisions made in CI/CD, which can help to further explore decision-making in CI/CD, as well as the difference compared to decision making in traditional software development~\cite{Li2019Decisions}. It is also interesting to further explore other data sources that communicate and make decisions in CI/CD and may partially answer the question why there are not many CI/CD-related decisions in issues. 

\subsection{Differences from Traditional Software Development}
We further compared our results (i.e., decision types made in CI/CD) with the study results by Li \emph{et al.} in our previous work \cite{Li2019Decisions} (i.e., decision types made in traditional open source development without using CI/CD) in this section.

From the perspective of decision categories, although there are differences in descriptions between our major decision categories and the classification of decisions in the Hibernate developer mailing list, they are similar in meaning and both basically cover the entire life cycle of development. The decision types we got in this work are more complete with one more category (i.e, \emph{Deployment Decision}).
In the decision subcategories, we did not identify \emph{Model Decision}, \emph{Pattern Decision}, \emph{Development Criteria Decision}, \emph{Implementation Decision}, and \emph{Annotation Decision}, but we got four more decision subcategories: \emph{Continuous Integration Decision}, \emph{Continuous Integration Testing Decision}, \emph{Traditional Deployment Decision}, and \emph{Continuous Deployment Decision}, which are largely related to CI/CD.

In terms of the percentages of various types of decisions, the largest proportion of decision type in traditional software development \cite{Li2019Decisions} is \emph{Design Decision} (42.6\%), followed by \emph{Requirement Decision} (31.6\%). The percentages of \emph{Management Decision} (10.1\%) and \emph{Construction Decision} (9.8\%) are roughly the same, and the percentage of \emph{Test Decision} (5.9\%) is the least.
However, the largest share of decision type in CI/CD is \emph{Requirement Decision} which accounts for 73.0\%, followed by \emph{Architecture Decision} (11.1\%) and \emph{Testing Decision} (9.2\%). For the remaining three decision types, each accounts for less than 6.0\%. 

Although these two studies used different data sources (i.e., issues in this work and developer mailing lists in \cite{Li2019Decisions}, respectively) to investigate the decisions made, \emph{Requirement Decision} and \emph{Design (Architecture) Decision} are dominant in both studies, and the percentages of \emph{Management Decision} and \emph{Construction Decision} in traditional development are higher than \emph{Management Decision} and \emph{Build Decision} in CI/CD, which is reasonable since the CI/CD pipeline handles most of management and build (construction) issues. \emph{Deployment Decision} only exists in CI/CD which is implied in the name of continuous deployment.

\section{Threats to Validity}\label{threat}
We discuss the potential threats to the validity of our results below by following the guideline in \cite{2012Experimentation}. Internal validity is not discussed because this aspect of validity is of concern when casual relationships are examined \cite{2012Experimentation}, and we did not investigate any causal relationships in our study.

\textbf{Construct validity} concerns generating the results of the study using the concept or theory behind the study \cite{2012Experimentation}. In our study, one potential threat to this validity comes from manual extraction and analysis of data. To migrate this threat, we randomly selected 50 issues from the dataset and discussed the criteria for data filtering, extraction, and analysis. Before the formal data extraction and analysis, the first and second authors had agreed on the criteria, and any uncertainty was discussed among the author to reach a consensus and eliminate personal bias. Another threat is that we did not consider whether the project is born with a CI/CD pipeline, and the statistics on the percentage of CI/CD decisions may be affected by the selection of the project that started using CI/CD late in the project life cycle.

\textbf{External validity} is concerned with to what extent it is possible to generalize the findings, and to what extent the findings are of interest to other people outside the investigated case \cite{2012Experimentation}. In this study, we analyzed the closed issues coming from a popular repository on GitHub, Budibase, which employs GitHub Actions to support CI/CD. One threat stems from the selection of the project and data source (issues), since it is possible that some other projects using other CI/CD tools (e.g., Travis CI) may also have relevant decisions in other data sources (e.g., pull requests). So our selected repository and data source may not be representative to all the CI/CD projects and data sources. To partially mitigate this threat, we conducted pilot project search with a set of criteria, and we also plan to include projects supported by other CI/CD tools with diverse data sources in our next step. Another threat is caused by the fact that we targeted one project only, and we will cover more repositories to migrate this threat.

\textbf{Reliability} is concerned with to what extent the data and the analysis are dependent on the specific researchers \cite{2012Experimentation}. One threat could be that the first author completed all the data extraction and analysis, and discussed the uncertain data with the second author to reach an agreement, which may lead to bias during data extraction and analysis. To alleviate this threat, we detailed the research process in Section \ref{research}, and the dataset and analysis results from the study have been made available online \cite{replpack}. With the measures stated above, we are confident that the study results are relatively reliable.

\section{Conclusions and Future Work}\label{conc}

As a set of established software quality assurance and development practices, continuous integration and continuous delivery are of interest to many researchers and practitioners, while the decisions made in CI/CD are not well understood. We conducted an exploratory study to obtain the types of decisions made in CI/CD using the closed issues (1,168) of Budibase as our dataset. After data filtering and extraction, we got 370 issues that contain decisions out of the 1,168 issues for further analysis. The findings of this study are the following: (1) We got 6 main categories and 11 subcategories of decisions made in CI/CD, in which \emph{Requirement Decision} has the highest percentage (67.6\%), with only a small amount of decisions (e.g., \emph{Continuous Deployment Decision}) related to CI/CD. (2) The types of decisions made in CI/CD we obtained in this study have certain common points with the types of decisions made in traditional software development, while \emph{Deployment Decision} only exists in CI/CD. (3) Practitioners are encouraged to put more effort on the issues and making decisions related to CI/CD, and researchers can extend the dataset and decision types got from this study by exploring projects employing other CI/CD tools and diverse data resources.

In the next step, we plan to extend this work on studying decisions made in CI/CD with a larger dataset from more repositories and diverse sources, and using complementary research methods (e.g., questionnaire, interview, and focus group). We also intend to take a deeper look at various aspects of decisions in CI/CD, including: (1) software artifacts involved in CI/CD-related decisions; (2) the rationale behind CI/CD decision-making; and (3) the impact of CI/CD-related decisions in development.


\balance
\bibliographystyle{IEEEtran}
\bibliography{References}

\end{document}